
\def\title{Structure of the space of states in RSOS models}
\def\author{Michio Jimbo${}^a$,
Tetsuji Miwa${}^b$ and Yasuhiro Ohta${}^c$\cr
{}\cr
${}^a$
Department of Mathematics,Faculty of Science,\cr
Kyoto University, Kyoto 606, Japan \cr
{}\cr
${}^b$
Research Institute for Mathematical Sciences,\cr
Kyoto University, Kyoto 606, Japan\cr
{}\cr
${}^c$
Faculty of Engineering, \cr
Hiroshima University, Higashi Hiroshima, Hiroshima 724 , Japan \cr
}
\par
\def\rhead{Space of states in RSOS models}
\def\lhead{M. Jimbo et al.}

\hsize=5.3truein
\vsize=7.8truein
\baselineskip=10pt
\font\twelvebf=cmbx12
\nopagenumbers
\headline={\ifnum\pageno=1 \hss\sl RIMS\hss \else \hdline\fi}
\def\hdline{\ifodd\pageno\rightheadline \else\leftheadline\fi}
\def\rightheadline{\tenrm\hfil{\it \rhead}\hfil\folio}
\def\leftheadline{\tenrm\folio\hfil{\it \lhead}\hfil}
\voffset=2\baselineskip
\vglue 2cm
%
%
\centerline{\twelvebf
\vbox{
\halign{\hfil # \hfil\cr
\title\crcr}}}
%
%
\bigskip
%
%
\centerline{\tenrm
\vbox{
\halign{\hfil#\hfil\cr
\author\crcr}}}
%
%
\vglue 1cm
%
%
%
\def\address #1
{\vglue 0.5cm
\halign{\quad\it##\hfil\cr
#1\crcr}}
%
\vskip 1.5cm

\par
{\narrower\bigskip{\noindent\bf Abstract.\quad}
The restricted solid-on-solid models in the anti-ferromagnetic
regime is studied in the framework of quantum affine algebras.
Following the line developed recently for vertex models,
a representation theoretical picture is presented for
the structure of the space of states.
The local operators and the creation/annihilation operators of quasi-particles
are defined using vertex operators, and their
commutation relations are calculated.
\bigskip}
\par

\message{Cross-reference macros, B. Davies, version 6 Aug 1992.}

\catcode`@=11


\newif\if@xrf\@xrffalse   
\def\l@bel #1 #2 #3>>{\expandafter\gdef\csname @@#1#2\endcsname{#3}}
\immediate\newread\xrffile
\def\xrf@n#1#2{\expandafter\expandafter\expandafter
\csname immediate\endcsname\csname #1\endcsname\xrffile#2}
\def\xrf@@n{\if@xrf\relax\else%
  \expandafter\xrf@n{openin}{ = \jobname.xrf}\relax%
  \ifeof\xrffile%
    \message{ no file \jobname.xrf - run again for correct forward references
}%
  \else%
    \expandafter\xrf@n{closein}{}\relax%
    \catcode`@=11 \input\jobname.xrf \catcode`@=12%
  \fi\global\@xrftrue%
  \expandafter\expandafter\csname immediate\endcsname%
  \csname  newwrite\endcsname\xrffile%
  \expandafter\xrf@n{openout}{ = \jobname.xrf}\relax\fi}


\newcount\t@g

\def\order#1{%
  \expandafter\expandafter\csname newcount\endcsname
  \csname t@g#1\endcsname\csname t@g#1\endcsname=0
  \expandafter\expandafter\csname newcount\endcsname
  \csname t@ghd#1\endcsname\csname t@ghd#1\endcsname=0

  \expandafter\def\csname #1\endcsname##1{\xrf@@n\csname n@#1\endcsname##1:>}

  \expandafter\def\csname n@#1\endcsname##1:##2>%
    {\def\n@xt{##1}\ifx\n@xt\empty%
     \expandafter\csname n@@#1\endcsname##1:##2:>
     \else\def\n@xt{##2}\ifx\n@xt\empty%
     \expandafter\csname n@@#1\endcsname\unp@ck##1 >:##2:>\else%
     \expandafter\csname n@@#1\endcsname\unp@ck##1 >:##2>\fi\fi}

  \expandafter\def\csname n@@#1\endcsname##1:##2:>%
    {\edef\t@g{\csname t@g#1\endcsname}\edef\t@@ghd{\csname t@ghd#1\endcsname}%
     \ifnum\t@@ghd=\t@ghd\else\global\t@@ghd=\number\t@ghd\global\t@g=0\fi%
     \ifunc@lled{@#1}{##1}\global\advance\t@g by 1%
       {\def\n@xt{##1}\ifx\n@xt\empty%
       \else\writ@new{#1}{##1}{\pret@g\t@ghead\number\t@g}\expandafter%
       \xdef\csname @#1##1\endcsname{\pret@g\t@ghead\number\t@g}\fi}%
       {\pret@g\t@ghead\number\t@g}%
     \else\def\n@xt{##1}%
       \w@rnmess#1,\n@xt>\csname @#1##1\endcsname%
     \fi##2}%

  \expandafter\def\csname ref#1\endcsname##1{\xrf@@n%
     \expandafter\each@rg\csname #1cite\endcsname{##1}}

  \expandafter\def\csname #1cite\endcsname##1:##2,%
    {\def\n@xt{##2}\ifx\n@xt\empty%
     \csname #1cit@\endcsname##1:##2:,\else%
       \csname #1cit@\endcsname##1:##2,\fi}

  \expandafter\def\csname #1cit@\endcsname##1:##2:,%
    {\def\n@xt{\unp@ck##1 >}\ifunc@lled{@#1}{\n@xt}%
      {\expandafter\ifx\csname @@#1\n@xt\endcsname\relax%
       \und@fmess#1,\n@xt>>>\n@xt<<%
       \else\csname @@#1\n@xt\endcsname##2\fi}%
     \else\csname @#1\n@xt\endcsname##2%
     \fi}}


\def\each@rg#1#2{{\let\thecsname=#1\expandafter\first@rg#2,\end,}}
\def\first@rg#1,{\callr@nge{#1}\apply@rg}
\def\apply@rg#1,{\ifx\end#1\let\n@xt=\relax%
\else,\callr@nge{#1}\let\n@xt=\apply@rg\fi\n@xt}

\def\callr@nge#1{\calldor@nge#1-\end-}
\def\callr@ngeat#1\end-{#1}
\def\calldor@nge#1-#2-{\ifx\end#2\thecsname#1:,%
  \else\thecsname#1:,\hbox{\rm--}\thecsname#2:,\callr@ngeat\fi}


\def\unp@ck#1 #2>{\unp@@k#1@> @>>}
\def\unp@@k#1 #2>>{\ifx#2@\@np@@k#1\else\@np@@k#1@> \unp@@k#2>>\fi}
\def\@np@@k#1#2#3>{\ifx#2@\@@np@@k#1>\else\@@np@@k#1>\@np@@k#2#3>\fi}
\def\@@np@@k#1>{\ifcat#1\alpha\expandafter\@@np@@@k\string#1>\else#1\fi}
\def\@@np@@@k#1#2>{@#2}


\def\writ@new#1#2#3{\xrf@@n\immediate\write\xrffile
  {\noexpand\l@bel #1 #2 {#3}>>}}


\def\ifunc@lled#1#2{\expandafter\ifx\csname #1#2\endcsname\relax}
\def\und@fmess#1#2,#3>{\ifx#1@%
  \message{ ** error - eqn label >>#3<< is undefined ** }\else
  \message{ ** error - #1#2 label >>#3<< is undefined ** }\fi}
\def\w@rnmess#1#2,#3>{\ifx#1@%
  \message{ Warning - duplicate eqn label >>#3<< }\else
  \message{ Warning - duplicate #1#2 label >>#3<< }\fi}


\def\t@ghead{}
\newcount\t@ghd\t@ghd=0
\def\taghead#1{\gdef\t@ghead{#1}\global\advance\t@ghd by 1}


\order{@qn}


\let\eqno@@=\eqno
\def\eqno(#1){\xrf@@n\eqno@@\hbox{{\rm(}$\@qn{#1}${\rm)}}}

\let\leqno@@=\leqno
\def\leqno(#1){\xrf@@n\leqno@@\hbox{{\rm(}$\@qn{#1}${\rm)}}}

\def\refeq#1{\xrf@@n{{\rm(}$\ref@qn{#1}${\rm)}}}


\def\eqalignno#1{\xrf@@n\displ@y \tabskip=\centering
  \halign to\displaywidth{\hfil$\displaystyle{##}$\tabskip=0pt
   &$\displaystyle{{}##}$\hfil\tabskip=\centering
   &\llap{$\eqaln@##$}\tabskip=0pt\crcr
   #1\crcr}}

\def\leqalignno#1{\xrf@@n\displ@y \tabskip=\centering
  \halign to\displaywidth{\hfil$\displaystyle{##}$\tabskip=0pt
   &$\displaystyle{{}##}$\hfil\tabskip=\centering
    &\kern-\displaywidth\rlap{$\eqaln@##$}\tabskip\displaywidth\crcr
   #1\crcr}}

\def\eqaln@#1#2{\relax\ifcat#1(\expandafter\eqno@\else\fi#1#2}
\def\eqno@(#1){\xrf@@n\hbox{{\rm(}$\@qn{#1}${\rm)}}}


\def\n@@me#1#2>{#2}
\def\numberby#1{\xrf@@n
  \ifx\s@ction\undefined\else
  \expandafter\let\csname\s@@ve\endcsname=\s@ction\fi
  \ifx\subs@ction\undefined\else
  \expandafter\let\csname\subs@@ve\endcsname=\subs@ction\fi
  \numb@rby#1,>#1>}
\def\numb@rby#1,#2>#3>{\def\n@xt{#1}\ifx\n@xt\empty\taghead{}\else
  \def\n@xt{#2}\ifx\n@xt\empty\n@by#3>\else\n@@by#3>\fi\fi}
\def\n@by#1>{\ifx\s@cno\undefined\expandafter\expandafter
  \csname newcount\endcsname\csname s@cno\endcsname
  \csname s@cno\endcsname=0\else\s@cno=0\fi
  \xdef\s@@ve{\expandafter\n@@me\string#1>}
  \let\s@ction=#1\def#1{\global\advance\s@cno by 1
  \taghead{\number\s@cno.}\s@ction}}
\def\n@@by#1,#2>{\ifx\s@cno\undefined\expandafter\expandafter
  \csname newcount\endcsname\csname s@cno\endcsname
  \csname s@cno\endcsname=0\else\s@cno=0\fi
  \ifx\subs@cst\undefined\expandafter\expandafter
  \csname newcount\endcsname\csname subs@cst\endcsname
  \csname subs@cst\endcsname=0\else\subs@cst=0\fi
  \ifx\subs@cno\undefined\expandafter\expandafter
  \csname newcount\endcsname\csname subs@cno\endcsname
  \csname subs@cno\endcsname=0\else\subs@cno=0\fi
  \xdef\s@@ve{\expandafter\n@@me\string#1>}
  \let\s@ction=#1\def#1{\global\advance\s@cno by 1
  \global\subs@cno=\subs@cst
  \taghead{\number\s@cno.}\s@ction}
  \xdef\subs@@ve{\expandafter\n@@me\string#2>}
  \let\subs@ction=#2\def#2{\global\advance\subs@cno by 1
  \taghead{\number\s@cno.\number\subs@cno.}\subs@ction}}


\def\numberfrom#1{\ifx\s@cno\undefined\else\n@mberfrom#1,>\fi}
\def\n@mberfrom#1,#2>{\def\n@xt{#2}%
  \ifx\n@xt\empty\n@@f#1>\else\n@@@f#1,#2>\fi}
\def\n@@f#1>{\s@cno=#1\advance\s@cno by -1}
\def\n@@@f#1,#2,>{\s@cno=#1\advance\s@cno by -1%
  \subs@cst=#2\advance\subs@cst by -1}


\def\pret@g{}
\def\prefixby#1{\gdef\pret@g{#1}}



\newcount\r@fcount\r@fcount=0
\newcount\r@fcurr
\newcount\r@fone
\newcount\r@ftwo
\newif\ifc@te\c@tefalse
\newif\ifr@feat

\def\refto#1{{\rm[}\def\s@p{}\refn@te#1>>\refc@te#1>>{\rm]}}

\def\refn@te#1>>{\refn@@te#1,>>}

\def\refn@@te#1,#2>>{\r@fnote{\expandafter\unp@ck\str@pbl#1 >> >}%
   \def\n@xt{#2}\ifx\n@xt\empty\else\refn@@te#2>>\fi}

\def\refc@te#1>>{\r@fcurr=0\r@featfalse\def\s@ve{}%
  {\loop\ifnum\r@fcurr<\r@fcount\advance\r@fcurr by 1\c@tefalse%
   \expandafter\refc@@te\number\r@fcurr>>#1,>>%
   \ifc@te\expandafter\refe@t\number\r@fcurr>>\fi\repeat\s@ve}}

\def\refc@@te#1>>#2,#3>>{\def\n@xt{\expandafter\unp@ck\str@pbl#2 >> >}%
   \expandafter\refc@@@te\csname r@f\n@xt\endcsname>>#1>>%
   \def\n@xt{#3}\ifx\n@xt\empty\else\refc@@te#1>>#3>>\fi}

\def\refc@@@te#1>>#2>>{\ifnum#2=#1\relax\c@tetrue\fi}

\def\refe@t#1>>{\ifr@feat\ifnum\r@fone=\r@ftwo\res@cond#1>>%
   \else\reth@rd#1>>\fi\else\r@feattrue\ref@rst#1>>\fi}

\def\ref@rst#1>>{\r@feattrue\r@fone=#1\r@ftwo=#1%
   \s@p\expandafter\relax\number\r@fone}%

\def\res@cond#1>>{\advance\r@ftwo by 1\def\n@xt{#1}%
   \expandafter\ifnum\n@xt=\number\r@ftwo%
   \edef\s@ve{,\expandafter\relax\number\r@ftwo}\else,\ref@rst#1>>\fi}%

\def\reth@rd#1>>{\advance\r@ftwo by 1\def\n@xt{#1}%
   \expandafter\ifnum\n@xt=\number\r@ftwo%
   \edef\s@ve{--\expandafter\relax\number\r@ftwo}\def\s@p{,}\else%
   \def\s@p{,}\s@ve\def\s@ve{}\ref@rst#1>>\fi}%

\def\r@fnote#1%
  {\ifunc@lled{r@f}{#1}\global\advance\r@fcount by 1%
   \expandafter\xdef\csname r@f#1\endcsname{\number\r@fcount}%
   \expandafter\gdef\csname r@ftext\number\r@fcount\endcsname%
   {\message{ Reference #1 to be supplied }%
   Reference $#1$ to be supplied\par}\fi}

\def\str@pbl#1 #2>>{#1#2}


\def\refis#1 #2\par{\def\n@xt{\unp@ck#1 >}\r@fis\n@xt>>#2>>}
\def\r@fis#1>>#2>>{\ifunc@lled{r@f}{#1}\else
   \expandafter\gdef\csname r@ftext\csname
r@f#1\endcsname\endcsname{#2\par}\fi}


\def\listreferences{\global\r@fcurr=0%
  {\loop\ifnum\r@fcurr<\r@fcount\global\advance\r@fcurr by 1%
   \numr@f\number\r@fcurr>>\csname r@ftext\number\r@fcurr\endcsname>>%
   \repeat}}

\def\numr@f#1>>#2>>{\vbox{\noindent\hang\hangindent=30truept%
   {\hbox to 30truept{\rm[#1]\hfill}}#2}\smallskip\par}


\def\printlabels{\global\@xrftrue\def\s@me##1{$##1$}
  \def\@qn##1{##1}\def\refeq##1{{\rm(}$##1${\rm)}}\refbylabel
  \def\numberby##1{\relax}\def\numberfrom##1{\relax}
  \def\listreferences{\relax}\def\referencefile{\relax}
  \def\order##1{\expandafter\let\csname ##1\endcsname=\s@me
                \expandafter\let\csname ref##1\endcsname=\s@me}}

\def\refbylabel{\def\refto##1{[$##1$]}%
  \def\refis##1 ##2\par{\numr@f$##1$>>##2>>}\def\numr@f##1>>##2>>%
  {\noindent\hang\hangindent=30truept{{\rm[##1]}\ }##2\par}}


\def\beginsection#1\par{\vskip0pt plus.3\vsize\penalty-250
  \vskip0pt plus -.3\vsize\bigskip\vskip\parskip
  \leftline{\bf#1}\nobreak\smallskip\noindent}

\catcode`@=12
%
%
\def\s{\sigma}
\def\o{\otimes}
\def\wo{\widehat{\otimes}}
\def\XXZ{{X\hskip-2pt X\hskip-2pt Z}}
%
%
\def\hom{\hbox{Hom}_{U'}}
\def\VN{V^{(N)}}
\def\vN{v^{(N)}}
\def\LN{L^{(N)}}
\def\BN{B^{(N)}}
\def\V1{V^{(1)}}
\def\qbinom#1#2{{#1}\atopwithdelims[]{#2}}
%
%
%
\def\hvarphi{\hat{\varphi}}
\def\C{{\bf C}}

\def\bR{\overline{R}}
\def\Q{{\bf Q}}
\def\Z{{\bf Z}}

\def\F{{\cal F}}

\def\la{\lambda}
\def\La{\Lambda}
\def\Lb{{\bar{\Lambda}}}
\def\etah{\hat{\eta}}
\def\xit{\tilde{\xi}}
\def\etat{\tilde{\eta}}
\def\Wb{\overline{W}}

\def\Vh{\widehat{V}}
\def\Lh{\widehat{L}}
\def\ep{\varepsilon}
\def\mod{\hbox{mod}}
\def\Hom{\mathop{\rm Hom}}
\def\End{\mathop{\rm End}}
\def\id{\hbox{id}}
\def\tr{\hbox{tr}}

\def\slth{\widehat{\goth{sl}}_2\hskip 1pt}
\def\uq{U_q(\goth{g})}

\def\uqa{U_q\bigl(\slth\bigr)}
\def\goto#1{{\buildrel #1 \over \longrightarrow}}
\def\br#1{\langle #1 \rangle}
\def\bra#1{\langle #1 |}
\def\ket#1{|#1\rangle}
\def\brak#1#2{\langle #1|#2\rangle}

\def\vac{|\hbox{vac}\rangle}

\def\BW#1#2#3#4#5{\left(\matrix{#1&#2 \cr #3&#4 \cr} \biggl|\,#5 \right)}
\def\wt{\hbox{wt}\,}
\def\Phit{\widetilde{\Phi}}

%
\font\germ=eufm10
\def\goth#1{\hbox{\germ #1}}
%
%
\def\Figure(#1|#2|#3)
{\midinsert
\vskip #2
\hsize 9cm
\raggedright
\noindent
{\bf Figure #1\quad} #3
\endinsert}
%
%
\def\Table #1. \size #2 \caption #3
{\midinsert
\vskip #2
\hsize 7cm
\raggedright
\noindent
{\bf Table #1.} #3
\endinsert}
%
\def\sectiontitle#1\par{\vskip0pt plus.1\vsize\penalty-250
 \vskip0pt plus-.1\vsize\bigskip\vskip\parskip
 \message{#1}\leftline{\bf#1}\nobreak\vglue 5pt}
\def\qed{\hbox{${\vcenter{\vbox{
    \hrule height 0.4pt\hbox{\vrule width 0.4pt height 6pt
    \kern5pt\vrule width 0.4pt}\hrule height 0.4pt}}}$}}
\def\subsec(#1|#2){\medskip\noindent{\it #1}\hskip8pt{\it #2}\quad}
\def\eq#1\endeq
{$$\eqalignno{#1}$$}
\def\leq#1\endeq
{$$\leqalignno{#1}$$}
%
%
\def\qbox#1{\quad\hbox{#1}\quad}
%
%

%
%

%
%

%
%

%
%

\def\Remark{\smallskip\noindent {\sl Remark.\quad}}
\def\Remarks#1{\smallskip\noindent {\sl Remark #1.\quad}}
%
%

%
%
\def\Definition#1.#2{\smallskip\noindent {\sl Definition #1.#2\quad}}
%
%
\def\subsec(#1|#2){\medskip\noindent#1\hskip8pt{\sl #2}\quad}
%
%
%
%
\def\abstract#1\endabstract{
\bigskip
\itemitem{{}}
{\bf Abstract.}
\quad
#1
\bigskip
}
%
%
%
\def\sec(#1){Sect.\hskip2pt#1}

\overfullrule=0pt
\order{prop}
\numberby{\beginsection}\numberfrom{0}
%
%
%
%
%
%

\beginsection \S 0. Introduction

Representation theory of quantized affine algebras
has brought several effective constructions of solutions to the
Yang-Baxter equation (see e.g. \refto{Ji,BS,FR}).
It is more than natural to expect that,
in solving the lattice models associated with the latter,
the quantized affine algebras would offer essential information.
The first step in this direction is to
understand the structure of the space of states, or the
`Hilbert space', of the model in terms of representation theory.
By space of states we mean
the one spanned by finite excitations over the ground states.

Such an approach was launched in \refto{DFJMN} for the XXZ model
\eq
&H_{\XXZ}=-{1\over 2}\sum_{k=-\infty}^\infty
\bigl( \s^x_{k+1}\s^x_k+\s^y_{k+1}\s^y_k
+\Delta\s^z_{k+1}\s^z_k \bigr) &(Ham)\cr
\endeq
in the anti-ferroelectric regime $\Delta<-1$.
The $\s^x_k,\s^y_k,\s^z_k$ are the Pauli matrices acting on the $k$-th
component of the tensor space $\cdots V\o V\o V\o \cdots$ ($V=\C^2$).
Suggested by the corner transfer matrix method
\refto{Baxbk, DJKMO, DJMO} and the crystal base theory \refto{(KMN)^2},
it was proposed \refto{DFJMN} that
the space of states is
the level $0$ module over the quantized affine algebra $\uqa$
\eq
& V(\La_i)\o V(\La_j)^*\simeq \hbox{Hom}_\C\left(V(\La_j),V(\La_i)\right).
&(Hilb)
\endeq
Here $V(\La_i)$ denotes the level one irreducible highest weight module
and $V(\La_i)^*$ is its dual module.
The choice of highest weight $\La_i$ (resp. $\La_j$) accomodates the boundary
condition to the left (resp. right) end of the lattice.
(To be precise, in \refeq{Hilb}
an appropriate completion of the tensor product is
necessary \refto{DFJMN}.)
Besides the manifest symmetry under $\uqa$,
an important feature of \refeq{Hilb} is that the
local structure of the lattice is automatically coded therein.
This statement is based on the fact
that after $q$-adic completion
the vertex operators provide
isomorphisms $V(\La_i)\rightarrow V(\La_{i+1})\o V$
(suffix $i$ to be read modulo $2$).
Iterating these one may `insert' the finite tensor product $V^{\o n}$
in \refeq{Hilb}
\eq
& V(\La_i)\o V(\La_j)^* \goto{\sim} V(\La_{i+n})\o V^{\o n} \o V(\La_j)^*,
\endeq
and vice versa.
This makes it possible to describe the local spin operators and their
correlation functions in terms of vertex operators \refto{JMMN}.
Analogous construction  is discussed in \refto{IIJMNT}
for the case of higher level representations of $\uqa$.

The aim of this note is to extend the above framework to
incorporate the restricted solid-on-solid (RSOS) models.
Suggested by the relevant results from the
corner transfer matrix \refto{ABF,DJKMO} and the
crystal base theory \refto{DJO},
we postulate that the `Hilbert space' in this case is
the space of $U'$-linear maps
\eq
&\hom \bigl(V(\xit)\o V(\etat), V(\xi)\o V(\eta) \bigr),
\endeq
where $U'$ is the subalgebra of $\uqa$ `with the scaling element
$d$ being dropped' \refto{DFJMN}.
Like in \refeq{Hilb}, the pair $(\xi,\eta)$ (resp. $(\xit,\etat)$)
corresponds to the choice of the boundary condition
to the left (resp. right) end of the lattice.
We shall show that the local structure of the lattice
and the creation/annihilation
of elementary excitations can be formulated using vertex operators.
For definiteness' sake we shall concentrate on the case of
$\uqa$ in this paper.
Evidently the same scheme applies to the general case.

It should be mentioned that closely related structures appear
in the literature on massive deformations of conformal field theory
\refto{Sm,Ber}.
Some of the formulas in this paper parallel those of the
deformation of the coset theory described in \refto{ABL}.
However the precise connection is yet to be uncovered.

The text is organized as follows.
To fix the notations we recall
in \sec(1) some relevant facts about the vertex operators.
In \sec(2), we define the space of states of the RSOS models.
In \sec(3) the transfer matrix, creation and annihilation operators are
introduced and their commutation relations are calculated.
In Appendix 1, restricting to the case of the Ising model
we will calculate the
eigenvectors of the transfer matrix by means of the vertex operators
to a few order in the $q$-series expansion.
Appendix 2 contains the explicit formulas for
the connection coefficients.

%

%
%
%
%

\beginsection \S 1. Vertex operators

\subsec(1.1|Notations)
Unless stated otherwise, we shall retain the notations of \refto{DFJMN},
such as
$\La_i$, $\alpha_i$, $h_i$ ($i=0,1$), $c$,
$d$, $\delta=\alpha_0+\alpha_1$,
$P=\Z\La_0\oplus\Z\La_1\oplus\Z\delta$,
$P^*=\Z h_0\oplus\Z h_1\oplus\Z d$
and $\rho=\La_0+\La_1$.
One change from \refto{DFJMN}
is that we normalize the invariant form $(~,~)$ on $P$
as $(\alpha_i,\alpha_i)=2$.
We identify $P^*$ with a subset of $P$ via $(~,~)$.
For $\la=m_0\La_0 +m_1\La_1 +n\delta$ we put
$\bar{\la}=m_1\bar{\La}_1$, $\bar{\La}_1=\La_1-\La_0$.
The quantized affine algebra $U=\uqa$
is defined by generators $e_i$, $f_i$, $t_i=q^{h_i}$ ($i=0,1$) and $q^d$
over $F=\Q(q)$,
and $U'=U'_q(\slth)$ signifies
the subalgebra generated by $e_i$, $f_i$ and $t_i$ ($i=0,1$).
The comultiplication $\Delta$ and the antipode $a$ are as in \refto{DFJMN}.

Set
$P_k=\{\la\in P \mid \br{h_i,\la}\in \Z_{\ge 0},~ \br{c,\la}=k~ \}$,
$P^0_k=\{\la\in P_k \mid \br{d,\la}=0\}$.
We denote by $V(\la)$ the irreducible module with highest weight $\la\in P_k$
and a fixed highest weight vector $u_\la\in V(\la)_\la$.
The weight space is denoted by $V(\la)_\nu$ ($\nu\in P$).

Let $\VN=\oplus_{j=0}^N F\vN_j$ denote the $(N+1)$-dimensional
$U'$-module given by ($\vN_{-1}=\vN_{N+1}=0$)
\eq
&e_1 \vN_j=[j]\vN_{j-1},\quad
 f_1 \vN_j=[N-j]\vN_{j+1},\quad
 t_1 \vN_j=q^{N-2j}\vN_j, \cr
&e_0=f_1,\quad f_0=e_1, \quad t_0=t_1^{-1} \quad \hbox{ on }\VN.
\endeq
In the case $N=1$ we shall also write $v_+=v^{(1)}_0$ and $v_-=v^{(1)}_1$.
Both $\VN$ and $\V1$ will play a role in the sequel.
Let $V$ be a finite dimensional $U'$-module.
As in \refto{(KMN)^2} its affinization means the $U$-module structure on
 $V_z=V\o F[z,z^{-1}]$ given as follows.
\eq
&e_i (v\o z^n)=(e_iv)\o z^{n+\delta_{i0}},
\quad (f_i v\o z^n)=(f_iv)\o z^{n-\delta_{i0}}, \cr
&t_i(v\o z^n)=(t_i v)\o z^n,\quad q^d(v\o z^n)= v\o (qz)^n.
&(affinization)
\endeq

For a $U$-(or $U'$-)module $M$ the dual module
$M^{*a^{\pm 1}}$ is defined to be
the restricted dual space $M^*=\oplus_\nu M_\nu^*$
endowed with the left module structure via $a^{\pm 1}$
($a$: antipode)
$$
\br{xu,v}=\br{u,{a^{\pm 1}(x)v}} \qbox{for }x\in U,\,u\in M^*,\,v\in M.
$$
The affinization $\left(V^{*a^{\pm 1}}\right)_z$ can be canonically
identified with the dual $\left(V_z\right)^{*a^{\pm 1}}$ in the above
sense via the pairing
$\br{u\o z^m,v\o z^n}=\br{u,v}\delta_{m+n\,0}$
($u\in V^*$, $v\in V$).
The modules $\VN_z$ are self-dual
in the sense that there exist the following isomorphisms of $U$-modules
\eq
&C_{\pm}^{(N)}~:~V_{zq^{\mp2}}^{(N)}~
\buildrel{\sim}\over\longrightarrow ~V_z^{(N)*a^{\pm1}},\qquad
C_{\pm}^{(N)} v_j^{(N)}= (-1)^j q^{-j(N-j\mp 1)}
{\qbinom{N}{j}}^{-1} v^{(N)*}_{N-j}.
&(dualV) \cr
\endeq
Here $\{v^{(N)*}_j \}$ signifies the dual base of $\{ \vN_j\}$
and \refeq{dualV} is to be extended $F[z,z^{-1}]$-linearly.
For $N=1$ \refeq{dualV} simplifies to $C^{(1)}_\pm v_+=v_-^*$,
$C^{(1)}_\pm v_-=-q^{\pm 1} v_+^*$.

\subsec(1.2|Vertex operators)
We recall below the properties of the vertex operators (VOs)
relevant to the subsequent discussions.
For more details see \refto{DJO,DFJMN}.

Fix positive integers $k$, $N$ such that $k\ge N$,
and let $\la,\mu\in P^0_k$.
We set $\Delta_\la=(\la,\la+2\rho)/2(k+2)$.
We shall use the following types of VOs:
\smallskip
\noindent{{\it Type I}}
\eq
&\Phi_\la^{\mu \VN}(z)=z^{\Delta_\mu-\Delta_\la}\Phit_\la^{\mu \VN}(z),
\qquad
\Phit_\la^{\mu \VN}(z): V(\la) \goto{} V(\mu)\wo \VN_z,
&(VOI:a) \cr
&\Phi_{\mu \VN}^\la(z)=z^{\Delta_\la-\Delta_\mu}\Phit_{\mu \VN}^\la(z),
\qquad
\Phit_{\mu \VN}^\la(z): V(\mu)\o \VN_z \goto{} V(\la)\wo F[z,z^{-1}], \cr
&&(VOI:b) \cr
\endeq
\noindent{{\it Type II}}
\eq
&\Phi_\la^{\VN \mu}(z)=z^{\Delta_\mu-\Delta_\la}\Phit_\la^{\VN \mu}(z),
\qquad
\Phit_\la^{\VN \mu}(z): V(\la) \goto{} \VN_z\wo V(\mu),
&(VOII:a)\cr
&\Phi_{\VN \mu}^\la(z)=z^{\Delta_\la-\Delta_\mu}\Phit_{\VN \mu}^\la(z),
\qquad
\Phit_{\VN \mu}^\la(z): \VN_z\o V(\mu) \goto{} V(\la)\wo F[z,z^{-1}]. \cr
&&(VOII:b) \cr
\endeq
\par
\noindent
For example \refeq{VOI:a} means a formal series of the form
\eq
&\Phit_\la^{\mu \VN}(z)=\sum \Phi_{j n} \o \vN_jz^{-n},
\quad
\Phi_{jn}:V(\la)_\nu~ \goto{}~V(\mu)_{\nu-\wt v_j+n\delta}
\quad \forall \nu, &(cmpt)
\endeq
such that it satisfies the intertwining relations
\eq
&\Phit_\la^{\mu \VN}(z)\circ x=\Delta(x)\circ\Phit_\la^{\mu \VN}(z)
\quad\forall x \in U.
\endeq
Here in the right hand side the action of $U$ on $V$ is in the sense of
\refeq{affinization}.
Likewise \refeq{VOI:b} means a formal series of
$F[z,z^{-1}]$-linear maps
\eq
&\Phit^\la_{\mu \VN}(z)(\cdot \o v_j)=\sum \Phi^*_{j n} \o z^{-n},
\quad
\Phi^*_{jn}:V(\mu)_\nu~ \goto{}~V(\la)_{\nu+\wt v_j+n\delta}\quad
\forall \nu, &(cmpt*)
\endeq
that enjoys similar intertwining properties.
We write
$\Phit_\la^{\mu \VN} =\Phit_\la^{\mu \VN}(1) =\sum \Phi_{j n} \o \vN_j$
to mean the corresponding
intertwiner of $U'$-modules $V(\la) \rightarrow \Vh(\mu)\o \VN$
where $\Vh(\mu)=\prod_\nu V(\mu)_\nu$. Similar convention is used for the
other types of VOs.

\subsec(1.3|Admissibility)
As in \sec(1.2) we let $k\ge N$.
The VOs \refeq{VOI:a}, \refeq{VOII:a}
are in one-to-one correspondence with the vectors
$v\in \VN$ such that \refto{DJO}
\eq
&e_i^{\br{h_i,\mu}+1}v=0~~(i=0,1),\quad
\wt v\equiv \la-\mu~~ \mod ~~\delta. &(VOexist)
\endeq
We say that a pair of weights $(\mu,\la)\in (P^0_k)^2$ is
$\VN$-{\it admissible}
if \refeq{VOexist} admits a nontrivial solution $v$ \refto{DJO}.
Since $\VN$ is multiplicity free, there exists at most one such $v$
up to proportionality.
If $\la=(k-a)\La_0+a\La_1$ and $\mu=(k-b)\La_0+b\La_1$, then a nontrivial
solution of \refeq{VOexist} exists if and only if
\eq
& a-b \in \{N,N-2,\cdots,-N\},\quad N\le a+b \le 2k-N. &(adm)
\endeq
In the special case $N=1$ this means that either $b=a+1$ with $a<k$, or
$b=a-1$ with $a>0$.

Choosing the scalar multiple of $v$ \refeq{VOexist} appropriately
we normalize the corresponding VOs as
\eq
&\Phit_\la^{\mu \VN}(z) u_\la = u_\mu \o \vN_j +\cdots,\qquad
\Phit_\la^{\VN \mu}(z) u_\la = \vN_j\o u_\mu +\cdots,
&(Normalize:a)\cr
&\Phit_{\mu \VN}^\la(z) \bigl(u_\mu \o \vN_j\bigr) = u_\la + \cdots,\qquad
\Phit_{\VN \mu}^\la(z) \bigl(\vN_j \o u_\mu \bigr) = u_\la + \cdots,
&(Normalize:b)
\cr
\endeq
where $j$ is such that $\la\equiv \mu+(N-2j)\Lb_1$ mod $\delta$.
In the first equation of \refeq{Normalize:a} the symbol
$\cdots$ indicates terms in $V(\mu)_\nu\o \VN$ with
$\nu\neq \mu$. Similarly for the other cases.

\Remarks{1}
For any $U'$-modules $M$, $N$, $\cdots$
we have canonical isomorphisms
\eq
&\hom (M_1\o M_2, N)~\goto{\sim}~\hom(M_2, M_1^{*a^{-1}}\o N), \cr
&\hom(M_1\o M_2, N)~\goto{\sim}~\hom(M_1, N\o M_2^{*a}). \cr
\endeq
Accordingly $\Phit_{\mu \VN}^\la(z)$ gives rise to the VO
\eq
&\Phit_\mu^{\la V^{(N)*a}}(z):V(\mu)\goto{}V(\la)\o V^{(N)*a},\cr
&\Phit_\mu^{\la V^{(N)*a}}(z)=\sum_{j,n} \Phi^*_{jn}\o v^{*(N)}_j
\endeq
where $\Phi^*_{jn}$ is given in \refeq{cmpt*}.
The normalization \refeq{Normalize:b} implies
$\Phit_\mu^{\la V^{(N)*a}}(z)u_\mu=u_\la\o v_j^{(N)*} +\cdots$.
Likewise $\Phit_{\VN \mu}^\la(z)$ gives rise to
$\Phit_\mu^{V^{(N)*a^{-1}}\la}(z)$.

\Remarks{2}
Let us call a (possibly infinite)
sequence $(\mu_j)_j$ of weights admissible if each $(\mu_j,\mu_{j+1})$ is.
Then, for each $n\ge 1$ and $\la,\la'$, the space of vertex operators
$V(\la)\rightarrow \Vh(\la')\o \bigl(\VN\bigr)^{\o n}$ has the basis
\eq
&\Phit^{\la' \VN}_{\mu_1}\circ \cdots \circ \Phit^{\mu_{n-1} \VN}_\la,
\qquad (\la',\mu_1,\cdots,\mu_{n-1},\la)\hbox{ : admissible}.
&(base_n)\cr
\endeq
\smallskip

\subsec(1.4|Crystals)
The representation $\VN$ has an upper crystal base $(\LN,\BN)$
given by $\LN=\oplus_{j=0}^N A v^{(N)}_j$,
$\BN=\{v_j~\mod~qL^{(N)} \}_{0\le j\le N}$
where $A=\{f(q)\in F\mid f(q) \hbox{ has } \break
\hbox{ no pole at }q=0 \}$.
Let $L(\la)$ denote the (upper) crystal lattice of $V(\la)$.
The type I VOs preserve the crystal lattice \refto{DJO} in the sense
$\Phit_\la^{\mu \VN}\left(L(\la)\right)\subset \Lh(\mu)\o \LN$,
where $\Lh(\mu)=\prod_\nu L(\mu)_\nu$.
It is known that
\eq
&\Phi_{\mu \VN}^\la(z)\circ \Phi_\la^{\mu \VN}(z)
=g_\la^{\mu} \times \id_{V(\la)}
&(gfac)
\endeq
with a scalar $g_\la^{\mu}=g_\la^{\mu}(k,N)$ \refto{DJO}.
The formula for $g_\la^{\mu}(k,1)$ is given in Appendix 2, \refeq{gfactor}.

The crystal $\BN$ is perfect of level $N$ in the sense of \refto{(KMN)^2}.
This implies that for each $\eta\in P^0_N$ of level $N$
there are unique weights $\etah$, $\etah'$
such that the pairs $(\eta,\etah)$, $(\etah',\eta)$ are admissible.
In our case we have $\etah=\etah'=\sigma(\eta)$ where
$\sigma$ denotes the involution
\eq
&\sigma\bigl((N-a)\La_0+a\La_1\bigr)=a\La_0+(N-a)\La_1. &(etahat)
\endeq
If $\la=\eta$ and $\mu=\sigma(\eta)$ have level $N$, then we have
in addition to \refeq{gfac}
\eq
&\Phi_\eta^{\sigma(\eta) \VN}(z)\circ \Phi_{\sigma(\eta) \VN}^\eta(z)
=g_\eta^{\sigma(\eta)} \times \id_{V(\etah)\o \VN}
\endeq
This shows that (after completion) the VO provides with an isomorphism
$\Phit_\eta^{\sigma(\eta) \VN}:V(\eta)\goto{}V(\sigma(\eta))\o \VN$.

\subsec(1.5|Connection coefficients)
Let
$\bR_{NN}(z_1/z_2)\in \End\bigl(\VN_{z_1}\o \VN_{z_2}\bigr)$
be the $R$ matrix
\eq
&P\bR_{NN}(z_1/z_2)\Delta(x)=\Delta(x)P\bR_{NN}(z_1/z_2)
\quad \forall x\in U,
\quad
Pu\o v=v\o u,
\endeq
normalized by
$\bR_{NN}(z_1/z_2)\vN_0\o \vN_0=\vN_0\o \vN_0$.
Let further $\la,\mu,\nu \in P^0_k$.
Then the following commutation
relation takes place among the VOs \refto{FR}:
\eq
&\bR_{NN}(z_1/z_2)\Phi_{\mu}^{\nu \VN_1}(z_1)\Phi_\la^{\mu \VN_2}(z_2) \cr
&\quad =\sum_{\mu'}
\Phi_{\mu'}^{\nu \VN_2}(z_2)\Phi_{\la}^{\mu' \VN_1}(z_1)
\times
W^N_k\BW {\la}{\mu}{\mu'}{\nu}{z_1/z_2} . &(connection)
\endeq
Here if
$\Phi_{\mu}^{\nu \VN}(z)=\sum_j \Phi_{\mu\, j}^{\nu \VN}(z)\o \vN_j$ then
we write
\eq
&\Phi_{\mu}^{\nu \VN_1}(z_1)\Phi_\la^{\mu \VN_2}(z_2)
=\sum_{jj'}\Phi_{\mu\,j}^{\nu \VN}(z_1)\Phi_{\la\,j'}^{\mu \VN}(z_2)
\o \vN_j\o \vN_{j'}, \cr
&\Phi_{\mu'}^{\nu \VN_2}(z_2)\Phi_{\la}^{\mu' \VN_1}(z_1)
=\sum_{jj'}
\Phi_{\mu'\,j'}^{\nu \VN}(z_2)\Phi_{\la\,j}^{\mu' \VN}(z_1)
\o \vN_j\o \vN_{j'}.\cr
\endeq
The connection coefficients
\eq
&W^N_k\BW{\la}{\mu}{\mu'}{\nu}
{z_1/z_2} &(Boltzmann)
\endeq
satisfy the Yang-Baxter equation in the face formulation.
Their explicit form for $N=1$ is included in the Appendix 2,
\refeq{Boltz,ABF}.

%
%
%
%
%
%

\beginsection \S 2. Space of states in the RSOS models

\subsec(2.1|RSOS models)
Let us recall the basic features of the solvable RSOS models
associated with $\uqa$ \refto{ABF, DJKMO}.
The model is labeled by a pair of positive integers $(l,N)$, $l=k-N$.
At each lattice site $(i,j)\in\Z^2$ one places a random variable
$\sigma_{i\,j}$
taking values in the set of local states
$S=\{0,1,\cdots,k\}$.
One imposes further a restriction
that for all $i,j$
the configurations on the adjacent lattice sites
$(\sigma_{i\,j},\sigma_{i+1\,j})$, $(\sigma_{i\,j+1},\sigma_{i\,j})$ be
admissible in the sense of \refeq{adm}.
We identify an element $a\in S$ with
the level $k$ dominant integral weight $\la_a=(k-a)\La_0+a\La_1\in P^0_k$.
The Boltzmann weight of the model is attached to each configuration
$(\la, \mu, \nu, \mu')$ on the NW, NE, SE, SW corners of
an elementary face, and is
given by the connection coefficients \refeq{Boltzmann}.

We shall consider exclusively the anti-ferromagnetic regime
(regime III of \refto{ABF}).
In the present parametrization this reads
\eq
&0<q^2<1, \qquad 1<z<q^{-2}. \cr
\endeq
Consider the limit of extreme order $q=0$.
In this limit only those configurations that are invariant under the
shift in the NE-SW direction survive. Hence they can be labeled by
configurations in the horizontal direction.
The ground states are given by
the sequences of states of the form $\cdots~b~c~b~c~b~c~\cdots$
with admissible $(b,c)\in S\times S$.
They correspond bijectively to a pair of dominant integral weights
$(\xi,\eta)\in P^0_l\times P^0_N$ of levels $l$ and $N$ respectively,
via $\la_b=\xi+\eta$, $\la_c=\xi+\sigma(\eta)$.
The excited states consist of admissible sequences (`restricted paths')
$\bigl(a^{(j)}\bigr)_{j\in\Z}$ satisfying
the boundary condition at the right (resp. left)
end of the lattice
specified by a ground state $(\xit,\etat)$ (resp. $(\xi,\eta)$).
This means that
$a^{(j)}=\xi+\sigma^j(\eta)$ for $j\ll 0$,
$a^{(j)}=\xit+\sigma^j(\etat)$ for $j\gg 0$.
(If we consider the periodic boundary condition then
only those states with $\xit=\xi$ and $\etat=\eta$ appear.)

We wish to find the eigenstates of the row transfer matrix
$T(z)$ at nonzero $q$.
We expect that
for $0<q^2<1$ these eigenstates can be expanded in powers of $q$, as
an infinite sum of the restricted paths obeying the same
boundary condition as the initial term $q=0$.
In the sequel we shall
present a mathematical definition of the space of eigenstates
consistent with the na\"{\i}ve picture described above.

\subsec(2.2|Space of states)
For $\xi\in P^0_l$ and $\eta\in P^0_N$, consider the tensor product
\eq
&V(\xi)\o V(\eta)=\oplus_{\la\in P^0_k} W(\xi\eta;\la),
\qquad
W(\xi\eta;\la)\simeq \Omega_{\xi\eta\la}\o V(\la),
&(isotypic)\cr
\endeq
where $\Omega_{\xi\eta\la}$ stands for the space of highest weight vectors
\eq
&\Omega_{\xi\eta\la}=\{v\in V(\xi)\o V(\eta)\mid
e_iv=0,~ t_iv=q^{\br{h_i,\la}}v~~\forall i\}. \cr
\endeq
We regard $\Omega_{\xi\eta\la}$ as a $U$-module
on which $U'$ acts trivially and
$q^d$ acts as $q^d\o q^d$ on $V(\xi)\o V(\eta)$.
It has been known by the corner transfer matrix method \refto{DJKMO},
and later systematized by the crystal base theory \refto{DJO}, that
one can label the base vectors of $\Omega_{\xi\eta\la}$
by restricted paths, by assigning suitable weights to the latter.

Now set
\eq
\F_{\xi\eta;\xit\etat}=&\hom(V(\xit)\o V(\etat),V(\xi)\o V(\eta)) \cr
=&\oplus_\la \hom(V(\la),V(\xi)\o V(\eta)) \o
\hom(V(\xit)\o V(\etat),V(\la))
&(FixBc)
\endeq
The second equality follows from the decomposition \refeq{isotypic}.
To be precise, we need a completion of this space in the $q$-adic sense
to accomodate the vertex operators which produce infinite sums.
To simplify the presentation, we will not discuss this point.
The symbols $\o$, etc. should be understood appropriately.

For each $n\ge 1$,  let $\eta^{(n)}=\sigma^n(\eta)$.
The iteration of \refeq{VOI} (with $z=1$) gives rise to
\eq
&\Phi^{(n)}:V(\eta)~ \goto{\sim}~ V(\eta^{(n)})\o \bigl(\VN\bigr)^{\o n}.
 &(VOntimes)
\endeq
This enables us to make an identification
\eq
&\hom\bigl(V(\la),V(\xi)\o V(\eta)\bigr) \cr
&\quad ~\goto{\sim}~
\hom\Bigl(V(\la),V(\xi)\o V(\eta^{(n)})\o \bigl(\VN\bigr)^{\o n}\Bigr) \cr
&\quad ~\goto{\sim}~
\oplus_{\la'}
\hom\Bigl(V(\la'),V(\xi) \o V(\eta^{(n)}) \Bigr) \o
\hom\Bigl(V(\la),V(\la') \o \bigl(\VN\bigr)^{\o n}\Bigr) \cr
&&(insert)\cr
\endeq
The first arrow is given by $\id\o\Phi^{(n)}$,
the second is given by the decomposition \refeq{isotypic}.

We take
\refeq{FixBc} to be the mathematical definition of the space
of states corresponding to the fixed boundary condition
$(\xi,\eta)$ (resp.  $(\xit,\etat)$) to the left (resp. right) half of the
horizontal line.
Because of the restriction condition in the model, horizontal configurations
between sites separated by $n$ steps are  admissible sequences
$(\la',\mu_1,\cdots,\mu_{n-1},\la)$.
In view of \refeq{base_n} we can identify them with the base vectors
of $\hom \bigl(V(\la),V(\la')\o (\VN)^{\o n}\bigr)$.
Therefore \refeq{insert} means that the na{\"\i}ve
local structure of the lattice can be embedded in the
space \refeq{FixBc}.

\subsec(2.3|Vacuum and inner product)
In the space $\F_{\xi\eta;\xi\eta}$ there is the unique
canonical element, i.e. the identity.
We call it the vacuum vector $\vac_{\xi\eta}=\id$.

For $f\in \F_{\xi\eta;\xi'\eta'}$
and
$g\in \F_{\xi'\eta';\xi\eta}$
we define a bilinear pairing
\eq
{\brak{f}{g}}_{\xi\eta}=&
{\tr_{V(\xi)\o V(\eta)}\bigl(q^{-2\rho}fg \bigr) \over
\tr_{V(\xi)\o V(\eta)}\bigl(q^{-2\rho} \bigr)}
=\sum_\la {\chi_\la \over \chi_\xi \chi_\eta}
\tr_{\Omega_{\xi\eta\la}}\bigl(q^{-2\rho}fg \bigr), &(pairing)\cr
\endeq
where
\eq
\chi_\la&=\tr_{V(\la)}\bigl(q^{-2\rho}\bigr)=q^{-(2\rho,\la)}(1+\cdots)~
\endeq
signifies the
prinicipally specialized character.

\Remark
The scaling element $d$ has a well-defined action on $V(\xi)\o V(\eta)$
and normalizes the action of $U'$: $[d,U']\subset U'$.
It follows that if $f\in \F_{\xi\eta;\xit\etat}$ then
$[d,f]\in \F_{\xi\eta;\xit\etat}$.
Hence $\F_{\xi\eta;\xit\etat}$ admits a $\Z$-grading by $d$.

%
%
%
%
%
\def\ptm{\phantom{*}}
\def\sid{{\scriptstyle id}}
\def\s{\sigma}

\beginsection \S 3. Particles

\subsec(3.1|Translation)
In the same spirit as in \refto{DFJMN} one may
introduce the left translation operator by a lattice unit
\eq
&T=T_{\eta\etat}^{\s(\eta)\s(\etat)}
{}~:~\F_{\xi\eta;\xit\etat}~\longrightarrow ~\F_{\xi\s(\eta);\xit\s(\etat)}.
\endeq
More generally we shall define $T(z)=T_{\eta\etat}^{\s(\eta)\s(\etat)}(z)$
which is related to $T$ by
\eq
&T(z)=z^{-d}T(1)z^d.&(boost)
\endeq
(Namely if $T=\sum T_n$ is the decomposition into the homogeneous
components with respect to the grading via $d$, then $T(z)=\sum T_nz^{-n}$.)
The image of a vector $f\in \F_{\xi\eta;\xit\etat}$ by $T(z)$,
$T(z)(f)\in \F_{\xi\s(\eta);\xit\s(\etat)}$,
is defined to be the composition
\eq
V(\xit)\o V(\s(\etat))
{}~\goto{\sid\o \Phi_1(z)}~
V(\xit)\o V(\etat) \o V^{(N)*a^{-1}}_z
{}~\goto{f\o \sid}~
&V(\xi)\o V(\eta) \o V^{(N)*a^{-1}}_z \cr
{}~\goto{\sid \o \Phi_2(z)}~
&V(\xi)\o V(\s(\eta)).
&(translation)
\endeq
We have set $\Phi_1(z)=\Phit_{\s(\etat)}^{\etat V^{(N)*a^{-1}}}(z)$
and $\Phi_2(z)=\Phi_{\eta V^{(N)*a^{-1}}}^{\s(\eta)}(z)$.
As in the vertex models it is natural to
identify $z^{-d}$ with the corner transfer matrix.
In view of \refeq{boost} we identify $T(z)$ with the row transfer matrix
of the RSOS model.

\Remark
In \refto{DFJMN,IIJMNT} the translation operator for the vertex models is
formulated as the composition
\eq
V(\la)\o V(\la')^{*a}
{}~\goto{\Phit_\la^{\s(\la)V}\o\sid}~
&V\bigl(\s(\la)\bigr)\o V \o V(\la')^{*a} \cr
{}~\goto{\sid\o \Phit_{V\la'}^{*\s(\la')} }~
&V\bigl(\s(\la)\bigr)\o V\bigl(\s(\la')\bigr)^{*a}
\endeq
where $V$ is perfect of level $k$, $\la, \la', \cdots \in P^0_k$, and
$\Phit_{V\la'}^{*\s(\la')}:V\o V(\la')^{*a}\rightarrow
V\bigl(\s(\la')\bigr)^{*a}$
is the transpose of
$\Phit_{\s(\la')}^{\la' V^{*a^{-1}} }
:V\bigl(\s(\la')\bigr) \rightarrow V(\la')\o V^{*a^{-1}}$.
Under the identification
$V(\la)\o V(\la')^{*a}\simeq \Hom\bigl(V(\la'),V(\la)\bigr)$,
this means that $f\in \Hom\bigl(V(\la'),V(\la)\bigr)$ is sent to
the composition
\eq
V\bigl(\s(\la')\bigr)
{}~ \goto{ \Phit_{\s(\la')}^{\la V^{*a^{-1}}} }~
V(\la')\o V^{*a^{-1}}
{}~\goto{f\o\sid}~
&V(\la)\o V^{*a^{-1}}
{}~\goto{ \Phit_{V(\la) V^{*a^{-1}} }^{\s(\la)} }~
V\bigl(\s(\la)\bigr).
\endeq
This can be compared with \refeq{translation}.

\subsec(3.2|Creation and annihilation)
Let $\xi\in P^0_l$, $\eta\in P^0_N$,
and let $\xi'\in P^0_l$, $\eta'\in P^0_N$ be such that
$(\xi',\xi)$, $(\eta',\eta)$ are
$\V1$-admissible.
We shall introduce $U$-linear operators
\eq
&\hvarphi^{*\xi'\eta'}_{\ptm \xi\eta}(z),~
\hvarphi^{\xi'\eta'}_{\xi\eta}(z)~:~
V(\xi)\o V(\eta) ~\goto{}~ V(\xi')\o V(\eta'). \cr
\endeq

By definition
$\hvarphi^{*\xi'\eta'}_{\ptm \xi\eta}(z)$
is the composition
\eq
V(\xi)\o V(\eta)
{}~~\goto{\Phi_\xi^{\xi' \V1}(z)\o \Phi_\eta^{V^{(1)*a^{-1}} \eta'}(z)}~~
&V(\xi')\o V^{(1)}_z\o V_z^{(1)*a^{-1}} \o V(\eta') \cr
\goto{\sid\o \br{~,~}\o\sid }~~
&V(\xi')\o V(\eta'). &(creation)
\endeq
Similarly
$\hvarphi^{\xi'\eta'}_{\xi\eta}(z)$
is the composition
\eq
V(\xi)\o V(\eta)
{}~~\goto{\Phi_\xi^{\xi'V^{(1)*a}}(z)\o\Phi_\eta^{\V1\eta'}(z)}~~
&V(\xi')\o V^{(1)*a}_z\o V^{(1)}_z \o V(\eta')  \cr
\goto{\sid\o\br{~,~}\o\sid}~~
&V(\xi')\o V(\eta'). &(annihilation)
\endeq
(Note that the canonical pairing
$\br{~,~}:V^{(1)}_z\o V_z^{(1)*a^{-1}} \rightarrow F$ or
$V^{(1)*a}_z\o V_z^{(1)} \rightarrow F$
is $U$-linear.)
In other words, we set
\eq
&\hvarphi^{*\xi'\eta'}_{\ptm \xi\eta}(z)
=\sum_{\ep=\pm}
\Phi_\xi^{\xi'\V1}(z)_\ep\o \Phi_\eta^{V^{(1)*a^{-1}}\eta'}(z)_\ep, \cr
&\hvarphi^{\xi'\eta'}_{\xi\eta}(z)
=\sum_{\ep=\pm}
\Phi_\xi^{\xi'V^{(1)*a}}(z)_\ep\o \Phi_\eta^{\V1 \eta'}(z)_\ep, \cr
\endeq
where
the components of the VOs are defined by
\eq
\Phi_\xi^{\xi'\V1}(z)&=\sum_\ep \Phi_\xi^{\xi'\V1}(z)_\ep\o v_\ep,
\quad
\Phi_\eta^{V^{(1)*a^{-1}}\eta'}(z)
=\sum_\ep v_\ep^*\o \Phi_\eta^{V^{(1)*a^{-1}}\eta'}(z)_\ep, \cr
\Phi_\xi^{\xi'V^{(1)*a}}(z)
&=\sum_\ep \Phi_\xi^{\xi'V^{(1)*a}}(z)_\ep\o v_\ep^*,
\quad
\Phi_\eta^{\V1 \eta'}(z)
=\sum_\ep v_\ep \o \Phi_\eta^{\V1 \eta'}(z)_\ep, \cr
\endeq
with $v_\ep^*$ being the dual base of $v_\ep$.
In view of the isomorphism \refeq{dualV} we have the relation
\eq
&\hvarphi^{\xi'\eta'}_{\xi\eta}(z)
=\hvarphi^{*\xi'\eta'}_{\ptm \xi\eta}(q^{-2}z)
\times
q^{2(\Delta_{\xi'}+\Delta_{\eta'}-\Delta_{\xi}-\Delta_{\eta})}
(-q)/(K_\xi^{\xi'} K_\eta^{\eta'}). \cr
\endeq
Here we have set
$K_\la^\mu=-q$ if $\mu=\la+\Lb_1$, $=1$ if $\mu=\la-\Lb_1$.

We now define the creation operator
$\varphi^{*\xi'\eta'}_{\ptm \xi\eta}(z)$
to be the map
\eq
&\F_{\xi\eta;\xit\etat} ~\goto{}~
\F_{\xi'\eta';\xit\etat},
\quad f \mapsto \hvarphi^{*\xi'\eta'}_{\ptm \xi\eta}(z)\circ f,
&(cre)
\endeq
where the quasi-momentum $z$ should lie on the unit circle
$|z|=1$.
Likewise we define the annihilation operator
$\varphi^{\xi'\eta'}_{\ptm \xi\eta}(z)$
to be the adjoint
(with respect to \refeq{pairing}) of
\eq
&\F_{\xi\eta;\xit\etat} ~\goto{}~
\F_{\xi\eta;\xit'\etat'},
\quad
f \mapsto f\circ \hvarphi^{\xit\etat}_{\xit'\etat'}(z).&(ann)
\endeq
Unlike the case of the vertex models \refto{DFJMN,IIJMNT},
the particles in the RSOS models
do not carry internal degees of freedom, or `spin'.

It should be mentioned that we do not have a mathematical proof
that \refeq{cre,ann} have well defined meaning on the space of states.
There are subtleties due to the fact that the type II VOs do not preserve
the crystal structure \refto{DFJMN}.
One may also ask whether one can define operators analogous to
\refeq{creation,annihilation}
by using other representations $V^{(l)}$ in place of $\V1$.
Though such operators do exist mathematically, we suspect that they
are not well defined in the above sense, cf. \refto{IIJMNT}.

\subsec(3.3|Momentum)
Using the definitions \refeq{translation,creation,annihilation}
it is possible to calculate the momentum of the particles created by
$\varphi^{*\xi'\eta'}_{\ptm\xi\eta}(z)$.
The following commutation relation can be deduced
from the $q$KZ equation (see \refto{IIJMNT}):
\eq
&\Phi_{\eta'}^{\s(\eta')\VN}(z_2) \circ
\Phi_\eta^{V^{(1)*a^{-1}}\eta'}(z_1)
=\tau(z_1/z_2)^{-1}
\Phi^{V^{(1)*a^{-1}}\s(\eta')}_{\s(\eta)}(z_1) \circ
\Phi_\eta^{\s(\eta)\VN}(z_2),
\endeq
where
\eq
&\tau(z)=
z^{-1/2}{\Theta_{q^4}(qz)\over \Theta_{q^4}(qz^{-1})}. \cr
\endeq
It then follows from the definitions of the operators
$T_{\eta\etat}^{\s(\eta)\s(\etat)}$ and
$\varphi^{*\xi'\eta'}_{\ptm\xi\eta}(z)$ that
\eq
&\widetilde{T}_{\eta' \etat}(z_2)
\circ \varphi^{*\xi'\eta'}_{\ptm\xi\eta}(z_1)
=\tau(z_1/z_2)^{-2}
\varphi^{*\xi'\eta'}_{\ptm\xi\eta}(z_1) \circ
\widetilde{T}_{\eta\etat}(z_2)
\endeq
where
$\widetilde{T}_{\eta\etat}(z)=
T^{\eta\etat}_{\s(\eta)\s(\etat)}(z)
T_{\eta\etat}^{\s(\eta)\s(\etat)}(z)$.
(Note that $\s$ is an involution.)
The formula for $\tau(z)$ is independent of the level $N$ and coincides with
the momentum of quasi-particles in the vertex model of spin $N/2$
\refto{IIJMNT}.
On the other hand, in both the vertex and RSOS models
the Hamiltonian can be obtained
by differentiating the row transfer matrix with respect to the
spectral parameter $z$.
We thus conclude that
the quasi-particles in the RSOS models carry the same energy and
momentum as in the vertex models.
However the particle structures are different because of the
admissibility condition in the RSOS models.

\subsec(3.4|Commutation relations)
The commutation relations of the operators
\refeq{creation,annihilation} can be deduced from those of
the vertex operators as given in the Appendix 2.
Restricting to $|z_i|=1$ we have
\eq
&\varphi_{\ptm\xi'\eta'}^{*\xi''\eta''}(z_1)
\varphi_{\ptm\xi\eta}^{*\xi'\eta'}(z_2) \cr
&\quad
=\sum_{\kappa,\zeta}
\varphi_{\ptm\zeta\phantom{''}\kappa}^{*\xi''\eta''}(z_2)
\varphi_{\ptm\xi\eta}^{*\zeta\kappa}(z_1)
\times
W^1_l\BW {\xi} {\xi'} {\zeta} {\xi''} {z_1/z_2}
W^{*1}_N\BW {\eta} {\eta'} {\kappa} {\eta''} {z_1/z_2} .
&(comm)
\endeq
The Boltzmann weights are given by \refeq{ABF}.
The $\varphi_{\xi\eta}^{\xi'\eta'}(z)$ enjoy the same commutation relations.
As for $\varphi_{\xi\eta}^{\xi'\eta'}(z)$ and
$\varphi_{\ptm\xi\eta}^{*\xi'\eta'}(z)$ we have
\eq
&\varphi_{\xi'\eta'}^{\xi''\eta''}(z_1)
\varphi_{\ptm\xi\eta}^{*\xi'\eta'}(z_2) \cr
&\quad
=\sum_{\kappa,\zeta}
\varphi_{\ptm\zeta\kappa}^{*\xi''\eta''}(z_2)
\varphi_{\xi\eta}^{\zeta\kappa}(z_1)
W^1_l\BW {\xi} {\xi'} {\zeta} {\xi''} {q^{-2}z_1/z_2}
W^{*1}_N\BW {\eta} {\eta'} {\kappa} {\eta''} {q^{-2}z_1/z_2} \cr
&\times (K_\xi^\zeta K_\eta^\kappa /K_{\xi'}^{\xi''}K_{\eta'}^{\eta''})
q^{2\Delta}
\cr
&+\delta_{\xi\xi''}\delta_{\eta\eta''}
g^{\xi'}_\xi(l,1) \tilde{g}^{\eta'}_\eta(N,1) \delta(z_1/z_2),
&(comm*)
\endeq
where $\Delta=\Delta_{\xi}+\Delta_{\xi''}+\Delta_{\eta}+\Delta_{\eta''}
-\Delta_{\xi'}-\Delta_{\zeta}-\Delta_{\eta'}-\Delta_{\kappa}$.
The delta function term comes from the simple pole of \refeq{comm*}
at $z_1=z_2$, and the forefactors
$g^{\xi'}_\xi(l,1)$,  $\tilde{g}^{\eta'}_\eta(N,1)$ are given in
Appendix 2, \refeq{gfactor,gtilde}.

The structure of the commutation relations \refeq{comm}
is precisely the same as that of the $S$-matrix for the
deformation of the coset conformal field theory \refto{ABL}.
In fact the latter is reproduced from \refeq{comm} in the limit
$q\rightarrow 1$.

\par
\bigskip\noindent
{\it Acknowledgement.}\quad
We wish to thank
A. Le Clair,
K. Miki,
T. Nakashima
and
A. Nakayashiki
for discussions.

%
\catcode`@=11
\newif\ifs@p
\def\refjl#1#2#3#4%
  {#1\def\l@st{#1}\ifx\l@st\empty\s@pfalse\else\s@ptrue\fi%
   \def\l@st{#2}\ifx\l@st\empty\else%
   \ifs@p, \fi{\frenchspacing\sl#2}\s@ptrue\fi%
   \def\l@st{#3}\ifx\l@st\empty\else\ifs@p, \fi{\bf#3}\s@ptrue\fi%
   \def\l@st{#4}\ifx\l@st\empty\else\ifs@p, \fi#4\s@ptrue\fi%
   \ifs@p.\fi\hfill\penalty-9000}
\def\refbk#1#2#3%
  {#1\def\l@st{#1}\ifx\l@st\empty\s@pfalse\else\s@ptrue\fi%
   \def\l@st{#2}\ifx\l@st\empty\else%
   \ifs@p, \fi{\frenchspacing\sl#2}\s@ptrue\fi%
   \def\l@st{#3}\ifx\l@st\empty\else\ifs@p, \fi#3\s@ptrue\fi%
   \ifs@p.\fi\hfill\penalty-9000}
\catcode`@=12
%
%

\def\CMP{Commun. Math. Phys.}

\def\IJMPA{Int. J. Mod. Phys. A}

\def\JSP{J. Stat. Phys.}

\def\NP{Nucl. Phys.}
\def\NPB{Nucl. Phys. B}

\def\RIMS{RIMS preprint}

\numberby{\beginsection}\prefixby{A}
\def\F{{\cal{F}}}
\def\ev{{\rm even}}
\def\od{{\rm odd}}
\def\Z{{\bf Z}}
\def\s{\sigma}
\def\La{\Lambda}
\def\la{\lambda}

\beginsection Appendix 1.

In this appendix, we will show how to get, in principle, the $q$-expansion
of the eigenvectors of the RSOS model by means of the VO method.
We will carry out the calculation to a few order in the Ising model,
and see that it actually gives the correct expansion. See \refto{FM,DFJMN}
for a similar calculation in the spin-1/2 XXZ model.

\subsec(A.1|Path expansions)
We call `restricted paths' simply `paths'. A path
$p=\bigl(p(l)\bigr)_{l\in\Z}$ is a sequence of integers, $p(l)\in S
=\{0,1,\ldots,k\}$, with the admissibility condition \refeq{adm}.
The `Hilbert space' of the RSOS model is formally
spanned by the paths: A `state' is written as
\eq
&|v\rangle=\sum_pc(p)|p\rangle,&(path)\cr
\endeq
where $c(p)$ is a formal power series in $q$.
We may obtain the set of states of this form
in two different ways:
\item{(i)}Perturbative diagonalization of the RSOS Hamiltonian.
\item{(ii)}Expansion of the states in $\F_{\xi\eta;\,\xit\etat}$.

At $q=0$ the RSOS Hamiltonian is diagonal \refto{DJO}. Therefore,
the paths $|p\rangle$ are eigenvectors.
The corresponding spectra are discrete,
although they are continuous for $q\not=0$.
Therefore, the spectra are infinitely degenerate at $q=0$.
In other words, the paths, which diagonalize the Hamiltonian at $q=0$,
are mixed states of the $q\not=0$ eigenvectors.
Although we expect the eigenvectors in the form \refeq{path},
we do not know how to start the expansion.
This is a difficulty in the approach (i).

For the lowest eigenvalue, this degeneracy is finite, and in fact,
it is resolved by specifying the boundary conditions. Choosing an admissible
pair $(b,c)$, we can calculate an expansion of the vacuum,
the lowest eigenvector of the RSOS Hamiltonian, for which the sum \refeq{path}
is restricted to the paths satisfying $p(2l)=\la_b$ and
$p(2l+1)=\la_c$ for large $|l|$.

We have another interpretation of \refeq{path}, i.e., (ii).
Consider \refeq{FixBc}, and take
a state $v$ in the subspace with a fixed $\lambda=\lambda_a$.
We interprete this $a\in S$ to be the value of the state
at the center of the one-dimensional lattice on which the RSOS Hamiltonian
is defined. Namely, such
$v$ is expanded in terms of $p$ satisfying $p(1)=a$.
Moreover, the component
$\F^-=\hom(V(\xit)\o V(\etat),V(\la))$
(resp. $\F^+=\hom(V(\la),V(\xi)\o V(\eta))$) represents
the right (resp. left) half of the restricted paths:
A vector $v^\pm$ in $\F^\pm$
has an expansion
\eq
&|v^\pm\rangle=\sum_{p^\pm}c(v^\pm;p^\pm)|p^\pm\rangle,&(leftpath)\cr
\endeq
where $p^+=\bigl(p^+(l)\bigr)_{l\ge1}$
(or $p^-=\bigl(p^-(l)\bigr)_{l\le1}$)
satisfies  $p^\pm(1)=\la$ and the appropriate
boundary condition for large $|l|$.

Let us consider $\F^+$.
The coefficients $c(p^+)$'s in \refeq{leftpath}
are determined from $v^+$ as follows.
Induce a map
\eq
\Phi^{\xi\eta}_{a(l),\ldots,a(1)}:
&\hom\bigl(V(\la_{a(1)}),V(\xi)\o V(\eta)\bigr)\rightarrow
\hom\bigl(V(\la_{a(l)}),V(\xi)\o V(\sigma^{l-1}\eta)\bigr) \cr
&&(Induce) \cr
\endeq
from \refeq{insert}: For $v^+$ in
\eq
&\hom\bigl(V(\la_{a(1)}),V(\xi)\o V(\eta)\bigr)\cr
&\simeq
\oplus_{\la'}
\hom\bigl(V(\la'),V(\xi)\o V(\sigma^{l-1}\eta)\bigr) \o
\hom\bigl(V(\la_{a(1)}),V(\la')\o V^{\o(l-1)}\bigr)\cr
\endeq
we have
\eq
&v^+=
\sum_{a(l),\ldots,a(2)}
\Phi^{\xi\eta}_{a(l),\ldots,a(1)}(v^+)
\otimes
\Phi^{\la_{a(l)}V}_{\la_{a(l-1)}}\circ\cdots
\circ\Phi^{\la_{a(2)}V}_{\la_{a(1)}}.
\cr
\endeq
We consider $\Phi^{\xi\eta}_{p(l),\ldots,p(1)}$.
If $l$ is large enough, we have $\la_{p^+(l)}=\xi+\sigma^{l-1}\eta$.
Therefore, we can define $c^{(l)}(v^+;p^+)$ to
be the coefficient of $u_\xi\otimes u_{\sigma^{l-1}\eta}$
considered as an element of
$\hom\bigl(V(\la_{p^+(l)}),V(\xi)\o V(\sigma^{l-1}\eta)\bigr)
\simeq\Omega_{\xi\,\sigma^{l-1}\eta\,\la_{a(l)}}
\subset V(\xi)\o V(\sigma^{l-1}\eta)$,
in the expansion of $\Phi^{\xi\eta}_{p^+(l),\ldots,p^+(1)}(v^+)$.
Then, we set
\eq
&c(p^+)=\lim_{l\rightarrow\infty}{c^{(l)}(v^+;p^+)
\over c^{(l)}(u_\xi\otimes u_\eta;p^+_{\xi\eta})},\cr
\endeq
where $p^+_{\xi\eta}=(\cdots\,a_1\,a_0\,a_1\,a_0)$
such that $\la_{a_0}=\xi+\eta$ and $\la_{a_1}=\xi+\sigma\eta$.

The right half is similar. Therefore, given a state $v$ in \refeq{FixBc},
we can expand it in the form \refeq{path}. We wish to check the
equality of (i) and (ii), e.g., for the vacuum state.
Or, a similar check is the following. We consider the CTM Hamiltonian.
As in \refto{FM}, we expect the equality between the following.
\item{(ia)}The perturbative diagonalization of the CTM Hamiltonian.
\item{(iia)}The path expansion of
$\hom\bigl(V(\la),V(\xi)\o V(\eta)\bigr)$ described above.

In the following section we report on our check for the lowest
eigenvector of the Ising CTM Hamiltonian versus the highest weight vector
$u_{\La_i}\otimes u_{\La_j}\in\hom\bigl(V(\La_i+\La_j),
V(\La_i)\o V(\La_j)\bigr)$.

\subsec(A.2|The lowest eigenvector of the Ising CTM Hamiltonian)
If $l=N=1$, the RSOS model reduces to two non-interacting Ising models.
The set $S$ is $\{0,1,2\}$. We identify $0$ with the $+$ Ising-spin
and $2$ with the $-$ Ising-spin. We write the Ising spin operator
by $\sigma^{x,y,z}_l$. The CTM Hamiltonians are
\eq
&B_\ev=-C_z\sum_{l\ge1}(2l-1)(\s^z_{2l+1}\s^z_{2l-1}-1)
+C_x\sum_{l\ge2}2(l-1)\s^x_{2l-1}+R_\ev,\cr
&B_\od=-C_z\sum_{l\ge1}2l(\s^z_{2l+2}\s^z_{2l}-1)
+C_x\sum_{l\ge1}(2l-1)\s^x_{2l}+R_\od,\cr
\endeq
where
\eq
&C_z={1\over4}-{\sum_{n\in\Z}(-)^nnq^{2n(2n-1)}\over
\sum_{n\in\Z}(-)^nq^{2n(2n-1)}}
={1\over4}+q^2+q^4+\cdots,\cr
&C_x=q{\sum_{n\in\Z}(-)^{n-1}nq^{4n(n-1)}\over
\sum_{n\in\Z}(-)^nq^{4n^2}}
=q+2q^5+\cdots,\cr
\endeq
and $R_{\ev/\od}$ is the renormalization constant to be determined.
The operator $B_{\ev}$ corresponds to the boundary condition
$(\xi,\eta)=(\La_0,\La_0)$ (or
$(\xi,\eta)=(\La_1,\La_1)$).
The operator $B_{\od}$ corresponds to the boundary condition
$(\xi,\eta)=(\La_0,\La_1)$ (or
$(\xi,\eta)=(\La_1,\La_0)$).
Without loss of generality, we restrict to the case $\xi=\La_0$.
It means at $q=0$ the lowest eigenvector reduces to a single path:
$(\cdots\,1\,2\,1\,2\,1\,2)$ (for $B_{\ev}$) or
$(\cdots\,2\,1\,2\,1\,2\,1)$ (for $B_{\od}$).
In the expansion of the lowest eigenvectors, the rightmost spins,
i.e., $2$ in the former or $1$ in the latter, are kept fixed.
The admissibility condition also freezes all the spins of the value $1$.
If we neglect the $1$'s, and only keep writing the Ising spins $-/+$ for $2/0$,
both cases are simply $(\cdots\,-\,-\,-)$. Let us use abbreviated symbols:
e.g.,
\eq
&|\phi\rangle=(\cdots\,-\,-\,-)\cr
&|1\rangle=(\cdots\,-\,-\,+)\cr
&|2\rangle=(\cdots\,-\,+\,-)\cr
&|2,1\rangle=(\cdots\,-\,+\,+)\cr
\endeq

We determine $R_{\ev/\od}$ and calculate the lowest eigenvector
$|vac\rangle=\sum c(p)|p\rangle$. We take
\eq
&R_\ev=\sum_{l\ge2}2(l-1)R_{2(l-1)}\quad\hbox{and}\quad
R_\od=\sum_{l\ge1}(2l-1)R_{2l-1},\cr
\endeq
where $R_l$ should be chosen so that
the lowest eigenvalue is zero.
The condition $B|vac\rangle=0$ gives us a way to calculate $c(p)$ and $R$.
Taking the coefficient of $|\phi\rangle$ in
$B|vac\rangle=\sum c(p)B|p\rangle$, we have
\eq
&\sum_{l\ge2}2(l-1)\bigr(C_xc(|l\rangle)+R_{2(l-1)}\bigl)=0,
\quad\hbox{for }B_\ev,\cr
&\sum_{l\ge1}(2l-1)\bigr(C_xc(|l\rangle)+R_{2l-1}\bigl)=0,
\quad\hbox{for }B_\od.\cr
\endeq
Therefore we choose
\eq
&R_{2(l-1)}=-C_xc(|l\rangle),\quad l\ge2,\cr
&R_{2l-1}=-C_xc(|l\rangle),\quad l\ge1.\cr
\endeq

We assume that $c(p)$ is expanded in $q$ as
\eq
&c(p)=\sum_{i\ge0}q^{m+2i}c_{2i}(p),\quad\hbox{for }
p=|j_m,j_{m-1},\ldots,j_1\rangle.\cr
\endeq
On this assumption, the coefficient $c(p)$ is successively and
consistently determined in the form of power series of $q$
by using the equation $B|vac\rangle=0$ expanded in $q$.
Let us give the path expansions of the lowest eigenvectors
with the coefficient of $|\phi\rangle$ normalized to $1$.
Keeping the paths $|j_m,j_{m-1},\ldots,j_1\rangle$ satisfying
$m\le3$ and the coefficients up to $q^5$, we have
\eq
&|vac\rangle_\ev=|\phi\rangle-(q-3q^3+6q^5)\sum_{m\ge2}|m\rangle
-q^5|2\rangle\cr
&\quad+(q^2-6q^4)\Bigl(\sum_{m>l\ge2}|m,l\rangle
+\sum_{m\ge2}|m+1,m\rangle\Bigr)+q^4\sum_{m\ge2}|m+2,m\rangle\cr
&\quad-(q^3-9q^5)\Bigl(\sum_{m>l>k\ge2}|m,l,k\rangle
+\sum_{m>l+1,l\ge2}|m,l+1,l\rangle+\sum_{m>l\ge2}|m+1,m,l\rangle\cr
&\quad+2\sum_{m\ge2}|m+2,m+1,m\rangle\Bigr)
+q^5\Bigl(\sum_{m\ge2}|m+2,m+1,m\rangle-\sum_{m>l+2,l\ge2}|m,l+2,l\rangle\cr
&\quad-\sum_{m>l\ge2}|m+2,m,l\rangle
-3\sum_{m\ge2}(|m+3,m+1,m\rangle+|m+3,m+2,m\rangle)\Bigr)+\cdots,\cr
&|vac\rangle_\od=|\phi\rangle-(q-3q^3+6q^5)\sum_{m\ge1}|m\rangle
-(q^3-3q^5)|1\rangle-q^5|2\rangle\cr
&\quad+(q^2-6q^4)\Bigl(\sum_{m>l\ge1}|m,l\rangle
+\sum_{m\ge1}|m+1,m\rangle\Bigr)
+q^4\Bigl(2|2,1\rangle+\sum_{m>1}|m,1\rangle+\sum_{m\ge1}|m+2,m\rangle\Bigr)\cr
&\quad-(q^3-9q^5)\Bigl(\sum_{m>l>k\ge1}|m,l,k\rangle
+\sum_{m>l+1,l\ge1}|m,l+1,l\rangle+\sum_{m>l\ge1}|m+1,m,l\rangle\cr
&\quad+2\sum_{m\ge1}|m+2,m+1,m\rangle\Bigr)
+q^5\Bigl(\sum_{m\ge1}|m+2,m+1,m\rangle
-\sum_{m>l+2,l\ge1}|m,l+2,l\rangle\cr
&\quad-\sum_{m>l\ge1}|m+2,m,l\rangle
-3\sum_{m\ge1}(|m+3,m+1,m\rangle+|m+3,m+2,m\rangle)\cr
&\quad-5|3,2,1\rangle-2\sum_{m>2}|m,2,1\rangle
-\sum_{m>l>1}|m,l,1\rangle-\sum_{m>1}|m+1,m,1\rangle\Bigr)+\cdots.\cr
\endeq

\subsec(A.3|Expansion of $u_{\La_0}\otimes u_{\La_i}$)
We use the following vectors in $V(\La_0)$.
\eq
&v_1=u_{\La_0},\quad
v_2=f_0v_1,\quad
v_3={1\over[2]}f_1v_2,\quad
v_4=f_0v_3,\cr
&v_5=f_1v_3,\quad
v_6={1\over[2]([3]-1)}([3]f_1v_4-f_0v_5),\quad
v_7={1\over[3]-1}(f_0v_5-f_1v_4).\cr
\endeq
By $v'_i\ (1\le i\le7)$ we mean the vectors in $V(\La_1)$
that are the images of $v_i\ (1\le i\le7)$ by the Dynkin diagram automorphism.

The first few highest weight vectors in $V(\La_0)\otimes V(\La_0)$
are as follows.
\eq
&w_1=v_1\otimes v_1\cr
&w_2=v_1\otimes v_2-qv_2\otimes v_1\cr
&w_3=v_1\otimes v_4-qv_2\otimes v_3+q^3v_3\otimes v_2-q^4v_4\otimes v_1\cr
&w_4=v_1\otimes v_7-qv_2\otimes v_5+q^3[2]v_3\otimes v_3-q^3v_5\otimes v_2
+q^6v_7\otimes v_1\cr
\endeq
Similarly, we have highest weight vectors in
$V(\La_0)\otimes V(\La_1)$.
\eq
&w'_1=v_1\otimes v'_1\cr
&w'_2=v_1\otimes v'_3-qv_2\otimes v'_2+q^3v_3\otimes v'_1\cr
&w'_3=v_1\otimes v'_6-qv_2\otimes v'_4+q^3v_3\otimes v'_3-q^4v_4\otimes v'_2
+q^6v_6\otimes v'_1\cr
\endeq
Their weights are
\eq
&\matrix{
w_1&w_2&w_3&w_4&w'_1&w'_2&w'_3\cr
2\La_0
&2\La_1-\delta
&2\La_1-2\delta
&2\La_0-2\delta
&\La_0+\La_1
&\La_0+\La_1-\delta
&\La_0+\La_1-2\delta\cr}\cr
\endeq
The vectors $w_1$ and $w_4$ are considered as vectors in
$\hom\bigl(V(2\La_0),V(\La_0)\o V(\La_0)\bigr)$,
$w_2$ and $w_3$ in
$\hom\bigl(V(2\La_1),V(\La_0)\o V(\La_0)\bigr)$,
and the rest in
$\hom\bigl(V(\La_0+\La_1),V(\La_0)\o V(\La_1)\bigr)$.

By using these vectors, we can write down a few terms
in the application of $\Phi_{a(2),a(1)}$ given in \refeq{Induce}:
\eq
&\Phi^{\La_0\La_0}_{12}(w_1)=
w'_1+{1\over[4]}w'_2+{[3]\over[8][2]}w'_3+\cdots,\cr
&\Phi^{\La_0\La_0}_{10}(w_2)=
w'_1-{[3]\over[4]}w'_2-{[5]\over[8][2]}w'_3+\cdots,\cr
&\Phi^{\La_0\La_0}_{10}(w_3)=
{1\over[2]}w'_1+{[5]\over[4][2]}w'_2-{[7][3]\over[8][2]^2}w'_3+\cdots,\cr
&\Phi^{\La_0\La_0}_{12}(w_4)=
-{1\over[4]}w'_1+{[5][3]\over[4]^2}w'_2-{[7][5]\over[8][4][2]}w'_3+\cdots,\cr
&\Phi^{\La_0\La_1}_{01}(w'_1)=
-{1\over[2]}w_2-{[3]\over[6][2]}w_3+\cdots,\cr
&\Phi^{\La_0\La_1}_{01}(w'_2)=
{[3]\over[2]^2}w_2-{[5][3]\over[6][2]^2}w_3+\cdots,\cr
&\Phi^{\La_0\La_1}_{01}(w'_3)=
{[5]\over[4][2]^2}w_2+{[7][3]^2\over[6][4][2]^2}w_3+\cdots,\cr
&\Phi^{\La_0\La_1}_{21}(w'_1)=
w_1-{[3]\over[6][4][2]}w_4+\cdots,\cr
&\Phi^{\La_0\La_1}_{21}(w'_2)=
{1\over[2]}w_1+{[5][3]^2\over[6][4][2]^2}w_4+\cdots,\cr
&\Phi^{\La_0\La_1}_{21}(w'_3)=
{[3]\over[4][2]}w_1-{[7][5][3]\over[6][4]^2[2]^2}w_4+\cdots.\cr
\endeq
Using these data we can calculate $c^{(l)}(v^+;p^+)$.
As an example let us show how the calculation goes for
$v^+=w_1$ and $p^+=p^+_{\La_0\La_0}$.
The operator $\Phi^{\xi\eta}_{p^+(l),\cdots,p^+(1)}$
is given by the composition
$\Phi^{\xi\,\sigma^l\eta}_{p^+(l),p^+(l-1)}\circ\cdots\circ
\Phi^{\xi\eta}_{p^+(2),p^+(1)}$. Applying these operators to $w_1$,
we get up to $q^6$
\eq
&\Phi^{\La_0\La_0}_{12}(w_1)=w'_1+(q^3-q^5)w'_2+q^6w'_3+\cdots,\cr
&\Phi^{\La_0\La_1}_{21}\circ\Phi^{\La_0\La_0}_{12}(w_1)=
(1+q^4-2q^6)w_1+q^5w_4+\cdots,\cr
&\Phi^{\La_0\La_0}_{12}\circ\Phi^{\La_0\La_1}_{21}\circ
\Phi^{\La_0\La_0}_{12}(w_1)=(1+q^4-2q^6)w'_1+q^3w'_2+\cdots,\cr
&\Phi^{\La_0\La_1}_{21}\circ\Phi^{\La_0\La_0}_{12}\circ
\Phi^{\La_0\La_1}_{21}\circ\Phi^{\La_0\La_0}_{12}(w_1)=
(1+2q^4-3q^6)w_1+q^5w_4+\cdots,\cr
&\Phi^{\La_0\La_0}_{12}\circ\Phi^{\La_0\La_1}_{21}\circ
\Phi^{\La_0\La_0}_{12}\circ\Phi^{\La_0\La_1}_{21}\circ
\Phi^{\La_0\La_0}_{12}(w_1)=(1+2q^4-3q^6)w'_1+q^3w'_2+\cdots.\cr
\endeq
The coefficient of $w_1$ (or $w'_1$) in the above expansion of
$\Phi_{p^+(l),\cdots,p^+(1)}(w_1)$ gives \hfil \break
$c^{(l)}(w_1;p^+_{\La_0\La_0})$,
that is, for $l\ge3$
\eq
&c^{(l)}(w_1;p^+_{\La_0\La_0})=1+Lq^4-(L+1)q^6+\cdots,\cr
\endeq
where $L=l/2-1$ for $l$:even (or $(l-1)/2$ for $l$:odd).
In a similar way, applying
$\Phi^{\xi\,\sigma^l\eta}_{p^+(l),p^+(l-1)}\circ\cdots\circ
\Phi^{\xi\eta}_{p^+(2),p^+(1)}$ to $w_1$ (or $w'_1$), we can calculate
$c^{(l)}(w_1;p^+)$ (or $c^{(l)}(w'_1;p^+)$) up to $q^5$ for the paths
$p^+=|m\rangle,|n,m\rangle$ and $|m+2,m+1,m\rangle$
for $n>m\ge2$ (or $n>m\ge1$). The results coincide with those obtained
by the perturbative calculation of $|vac\rangle$ in A.2.
Therefore it is strongly suggested that the equality holds between
the perturbative diagonalization of the CTM Hamiltonian and
the path expansion of the states in $\F^+$.

%
%
%
%

\beginsection Appendix 2. Connection coefficients

Here we include the explicit form of the connection coefficients
in the case $N=1$ \refto{FR}.
Fix $\la\in P^0_k$, and set
\eq
&p=q^{2(k+2)}, \quad s={1\over 2(k+2)},\quad \la_{\pm}=\la\pm \Lb_1,
\quad r_+=1-r_-,\quad r_-=\Delta_{\la_+}-\Delta_{\la_-},\cr
&(z;p)_\infty=\prod_{j\ge 0}(1-zp^j), \quad
(z;p,p')_\infty=\prod_{i,j\ge 0}(1-zp^i{p'}^j), \cr
&\Theta_p(z)=(z;p)_\infty(p/z;p)_\infty(p;p)_\infty,
\quad \Gamma_p(x)={(p;p)_\infty \over (p^x;p)_\infty}(1-p)^{1-x}, \cr
&\psi(z)={(q^4z;q^4,p)_\infty (z;q^4,p)_\infty
\over (q^2z;q^4,p)_\infty^2}.
\endeq
Let $V=\V1$ and let $\bR_{11}(z)=\bR(z)$ be as in \refto{DFJMN}, eq.(6.16).
Then
\eq
&\bR(z_1/z_2)\Phi_\mu^{\nu V_1}(z_1)\Phi_\la^{\mu V_2}(z_2) \cr
&\quad =\sum_{\mu'} \Phi_{\mu'}^{\nu V_2}(z_2)\Phi_\la^{\mu' V_1}(z_1)
W^1_k\BW{\la}{\mu}{\mu'}{\nu}{z_1/z_2},\cr
&\left(\bR(z_1/z_2)^t\right)^{-1}
\Phi_\mu^{V_1^{*a^{-1}}\nu}(z_1)\Phi_\la^{V_2^{*a^{-1}}\mu }(z_2) \cr
&\quad =\sum_{\mu'}
\Phi_{\mu'}^{V_2^{*a^{-1}}\nu}(z_2)\Phi_\la^{V_1^{*a^{-1}}\mu' }(z_1)
W^{*1}_k\BW{\la}{\mu}{\mu'}{\nu}{z_1/z_2},
\endeq
where
\eq
W^1_k\BW{\la}{\mu}{\mu'}{\nu}{z}
&=\Wb^1_k\BW{\la}{\mu}{\mu'}{\nu}{z}
\times {\psi(pz^{-1}) \over \psi(pz)}
z^{\Delta_\la+\Delta_\nu-\Delta_\mu-\Delta_{\mu'}}
&(Boltz:a)\cr
W^{*1}_k\BW{\la}{\mu}{\mu'}{\nu}{z}
&=\Wb^1_k\BW{\la}{\mu}{\mu'}{\nu}{z}
\times
{\psi(z^{-1}) \over \psi(z)}
z^{\Delta_\la+\Delta_\nu-\Delta_\mu-\Delta_{\mu'}}
&(Boltz:b)
\endeq
and
\eq
&\Wb^1_k\BW{\la}{\mu}{\mu}{\nu}{z}=1 \qquad
(\mu-\la=\nu-\mu=\pm {\bar \Lambda}_1), &(ABF:a)\cr
&\Wb^1_k\BW{\la}{\la_\pm}{\la_\pm}{\la}{z}=
{\Theta_p(q^2)\Theta_p(p^{r_\pm}z)
\over \Theta_p(p^{r_\pm})\Theta_p(q^2z)}z^{(1\pm 1)/2},
&(ABF:b)\cr
&\Wb^1_k\BW{\la}{\la_\mp}{\la_\pm}{\la}{z}=
q{\Gamma_p(r_\pm)\Gamma_p(r_\pm) \over \Gamma_p(2s+r_\pm)\Gamma_p(-2s+r_\pm)}
{\Theta_p(z) \over \Theta_p(q^2z)}. &(ABF:c)\cr
\endeq
Up to a `gauge' they coincide with the Boltzmann weights due to
Andrews-Baxter-Forrester \refto{ABF}.
For general $N$ the weights \refeq{Boltzmann}
are constructed from \refeq{ABF} through the fusion procedure \refto{DJKMO}.

The following formulas are used in \sec(3):
\eq
g^{\la_\pm}_\la(k,1)&=
\bra{u_\la} \Phit_{\la_\pm \V1}^\la(z)\Phit_\la^{\la_\pm \V1}(z)\ket{u_\la}\cr
&={(pq^2;q^4,p)_\infty^2 \over (p;q^4,p)_\infty (pq^4;q^4,p)_\infty}
{(p^{r_\pm}q^2;p)_\infty \over (p^{r_\pm};p)_\infty}, &(gfactor)\cr
\tilde{g}^{\la_\pm}_\la(k,1)&=
Res_{z_1=z_2}\bra{u_\la} \Phit_{\la_\pm}^{\V1 \la}(z_1)
\Phit_{\V1 \la}^{\la_\pm}(z_2)\ket{u_\la}d(z_1/z_2)\cr
&={(q^2;q^4,p)_\infty^2 \over (q^4;q^4,p)_\infty^2 (p;p)_\infty}
{(p^{r_\pm}q^2;p)_\infty \over (p^{r_\pm};p)_\infty}. &(gtilde)\cr
\endeq

\bigskip\noindent{\bf References}\medskip
\par

\refis{Ji} \refjl
{see e.g. Jimbo M, Introduction to the Yang-Baxter equation}
{\IJMPA}{4}{(1989) 3759-3777}

\refis{BS} \refjl
{Bazhanov V V and Stroganov Yu G,
Chiral Potts model as a descendant of the six vertex models}
{\JSP}{51}{(1990) 799--817}

\refis{FR} \refjl
{Frenkel I B and Reshetikhin N Yu,
Quantum affine algebras and holonomic difference equations}
{\CMP}{146}{(1992) 1--60}
\par

\refis{DFJMN} \refjl
{Davies B, Foda O, Jimbo M, Miwa T and Nakayashiki A,
Diagonalization of the XXZ Hamiltonian by vertex operators}
{\RIMS}{873}{(1992)}
\par

\refis{Baxbk} \refbk
{Baxter R J}
{Exactly solved models in statistical mechanics}
{Academic Press, London 1982}

\refis{DJKMO} \refjl
{Date E, Jimbo M, Kuniba A, Miwa T and Okado M,
Exactly solvable SOS models:
Local height probabilities and theta function identities}
{\NPB}{290}{[FS20] (1987) 231--273}

\refis{DJMO} \refjl
{Date, E., Jimbo, M., Kuniba, A., Miwa, T. and Okado, M,
One dimensional configuration sums in vertex models and affine Lie
algebra characters}
{Lett. Math. Phys.} {17} {(1989) 69--77}

\refis{(KMN)^2} \refjl
{Kang S-J, Kashiwara M, Misra K, Miwa T, Nakashima T and Nakayashiki A,
Affine crystals and vertex models}
{\IJMPA}{7}{Suppl. 1A (1992) 449--484}

\refis{JMMN} \refjl
{Jimbo M, Miki K, Miwa T and Nakayashiki A,
Correlation functions of the XXZ model for $\Delta<-1$}
{\RIMS}{877}{(1992)}
\par

\refis{IIJMNT} \refjl
{Idzumi M, Iohara K, Jimbo M, Miwa T, Nakashima T and Tokihiro T,
Quantum affine symmetry in vertex models}
{\RIMS}{}{(1992) }
\par

\refis{ABF} \refjl
{Andrews G E, Baxter R J and Forrester P J,
Eight-vertex $SOS$ model and generalized Rogers-Ramanujan-type identities}
{\JSP}{35}{(1984) 193-266}

\refis{DJO} \refjl
{Date E, Jimbo M and Okado M,
Crystal base and $q$ vertex operators}
{Osaka Univ. Math. Sci. preprint}{1}{(1991)}
\par

\refis{Sm} \refjl
{Smirnov F A,
Dynamical symmetries of massive integrable models}
{\IJMPA}{7}{Suppl. 1B (1992) 813--837, 839--858}

\refis{Ber} \refjl
{Bernard D,
Hidden Yangians in 2D massive current algebras}
{\CMP}{137}{(1991) 191--208}

\refis{ABL} \refjl
{Ahn C, Bernard D and LeClair A,
Fractional supersymmetries in perturbed coset CFT's
and integrable soliton theory}
{\NP}{B340}{(1990) 721}

\refis{FM} \refjl
{Foda O and Miwa T,
Corner transfer matrices and quantum affine algebras}
{\IJMPA}{7}{Suppl. 1A (1992) 1--53}
\par

\listreferences
\par

\def\uq{U_q\bigl(\widehat{\goth{sl}}\hskip2pt(n)\bigr)}

\refis{MM} \refjl
{Misra K C and Miwa T,
Crystal base for the basic representation of $\uq$}
{\CMP}{134}{(1990) 79--88}

\refis{JMMO} \refjl
{Jimbo M, Misra K C, Miwa T and Okado M,
Combinatorics of representations of $\uq$ at $q=0$}
{\CMP}{136}{(1991) 543--566}

\refis{Pas} \refjl
{Pasquier V,
Etiology of IRF models}
{\CMP}{118}{(1988) 335-364}

\end